\documentstyle[12pt,aaspp4]{article}

\def \ref {\noindent\hangindent=1.0in\hangafter=1}
\def \cl {\centerline}

\def\ltsima{$\; \buildrel < \over \sim \;$}
\def\simlt{\lower.5ex\hbox{\ltsima}} % < over ~
\def\gtsima{$\; \buildrel > \over \sim \;$}
\def\simgt{\lower.5ex\hbox{\gtsima}} % > over ~

\begin{document}
%\today
\title{Multiwavelength Observations of a Dramatic High Energy Flare
in the Blazar 3C~279 }
\author{A. E. Wehrle\altaffilmark{1}, 
E. Pian\altaffilmark{2,3}, 
C. M. Urry\altaffilmark{2}, 
L. Maraschi\altaffilmark{4}, 
G. Ghisellini\altaffilmark{5}, 
R. C. Hartman\altaffilmark{6},
G. M. Madejski\altaffilmark{6}, 
F. Makino\altaffilmark{7}, 
A. P. Marscher\altaffilmark{8}, 
I. M. McHardy\altaffilmark{9}, 
S. J. Wagner\altaffilmark{10},
J. R. Webb\altaffilmark{11,12}, 
G. S. Aldering\altaffilmark{13}, 
M. F. Aller\altaffilmark{14}, 
H. D. Aller\altaffilmark{14}, 
D. E. Backman\altaffilmark{15},
T. J. Balonek\altaffilmark{16}, 
P. Boltwood\altaffilmark{17},
J. Bonnell\altaffilmark{6}, 
J. Caplinger\altaffilmark{6}, 
A. Celotti\altaffilmark{18}, 
W. Collmar\altaffilmark{19},
J. Dalton\altaffilmark{15},
A. Drucker\altaffilmark{15}, 
R. Falomo\altaffilmark{20}, 
C. E. Fichtel\altaffilmark{6}, 
W. Freudling\altaffilmark{21}, 
W. K. Gear\altaffilmark{22},
N. Gonzalez-Perez\altaffilmark{23}, 
P. Hall\altaffilmark{24}, 
H. Inoue\altaffilmark{7},
W. N. Johnson\altaffilmark{25}, 
D. Kazanas\altaffilmark{6},
M. R. Kidger\altaffilmark{23},
T. Kii\altaffilmark{7},
R. I. Kollgaard\altaffilmark{26},
Y. Kondo\altaffilmark{6}, 
J. Kurfess\altaffilmark{25},
A. J. Lawson\altaffilmark{9},
Y. C. Lin\altaffilmark{27},
B. McCollum\altaffilmark{6}, 
K. McNaron-Brown\altaffilmark{25}, 
F. Nagase\altaffilmark{7},
A. D. Nair\altaffilmark{28}, 
S. Penton\altaffilmark{29}, 
J. E. Pesce\altaffilmark{2,30}, 
M. Pohl\altaffilmark{19},
C. M. Raiteri\altaffilmark{31},
M. Renda\altaffilmark{15},
E. I. Robson\altaffilmark{22,32}, 
R. M. Sambruna\altaffilmark{6,30},
A. F. Schirmer\altaffilmark{16}, 
C. Shrader\altaffilmark{6}, 
M. Sikora\altaffilmark{33},
A. Sillanp\"a\"a\altaffilmark{34}, 
P. S. Smith\altaffilmark{35},
J. A. Stevens\altaffilmark{32}, 
J. Stocke\altaffilmark{29}, 
L. O. Takalo\altaffilmark{34}, 
H. Ter\"asranta\altaffilmark{36}, 
D. J. Thompson\altaffilmark{6}, 
R. Thompson\altaffilmark{6},
M. Tornikoski\altaffilmark{36}, 
G. Tosti\altaffilmark{37}, 
P. Turcotte\altaffilmark{15}, 
A. Treves\altaffilmark{38},
S. C. Unwin\altaffilmark{39}, 
E. Valtaoja\altaffilmark{36},
M. Villata\altaffilmark{31}, 
W. Xu\altaffilmark{1}, 
A. Yamashita\altaffilmark{7},
A. Zook\altaffilmark{40} 
}

%  AFFILIATIONS:
\altaffiltext{1}{Infrared Processing Analysis Center, MC 100-22, Jet
Propulsion Laboratory and California Institute of Technology,
Pasadena, CA 91125}
\altaffiltext{2}{Space Telescope Science Institute, 3700 San Martin Drive, 
Baltimore, MD 21218}
\altaffiltext{3}{ITESRE-CNR, Via Gobetti 101, I-40129 Bologna, Italy}
\altaffiltext{4}{Osservatorio Astronomico di Brera, Via Brera 28, I-20121
Milan, Italy}
\altaffiltext{5}{Osservatorio Astronomico di Brera, Via Bianchi 46, I-22055
Merate (Lecco), Italy}
\altaffiltext{6}{NASA/Goddard Space Flight Center, Greenbelt, MD 20771}
\altaffiltext{7}{ISAS, 3-1-1, Yoshinodai, Sagamihara, Kanagawa 229, Japan}
\altaffiltext{8}{Department of Astronomy, Boston University, 725 Commonwealth 
Avenue, Boston, MA 02215}
\altaffiltext{9}{Department of Physics, University of Southampton, Southampton 
SO9 5NH, United Kingdom}
\altaffiltext{10}{Landessternwarte, Heidelberg-K\"onigsstuhl, D-69117
Heidelberg, Germany}
\altaffiltext{11}{Department of Physics, Florida International University, 
University Park, Miami, FL 33199}
\altaffiltext{12}{SARA Observatory, KPNO/NOAO, 950 North Cherry Avenue,
P.O. Box 26732, Tucson, AZ 85726}
\altaffiltext{13}{University of Minnesota, Department of Astronomy, 116 Church
St., SE Minneapolis, MN 55455}
\altaffiltext{14}{University of Michigan, Physics and Astronomy, 817
Dennison Building, Ann Arbor, MI 48109}
\altaffiltext{15}{Franklin \& Marshall College, Physics \& Astronomy 
Department, P.O. Box 3003, Lancaster, PA 17604-3003}
\altaffiltext{16}{Colgate University, Department of Physics \& Astronomy,
13 Oak Dr., Hamilton, NY 13346-1398}
\altaffiltext{17}{1655 Main St., Stittsville, Ontario, Canada K2S 1N6}
\altaffiltext{18}{SISSA/ISAS, Via Beirut 2-4, I-34014 Miramare-Grignano
(Trieste), Italy}
\altaffiltext{19}{Max Planck-Institut f\"ur Extraterrestrische Physik,
Giessenbachstrasse, D-85740 Garching bei M\"unchen, Germany}
\altaffiltext{20}{Osservatorio Astronomico di Padova, Via Osservatorio 5,
I-35122 Padova, Italy}
\altaffiltext{21}{Space Telescope European Coordinating Facility, 
Karl-Schwarzschild Strasse 2, D-85748 Garching bei M\"unchen, Germany}
\altaffiltext{22}{Centre for Astrophysics, University of Central Lancashire,
Preston, PR1 2HE, United Kingdom}
\altaffiltext{23}{Instituto de Astrof\'isica de Canarias, E-38200, La
Laguna, Tenerife, Spain}
\altaffiltext{24}{Steward Observatory, University of Arizona, Tucson AZ 85721}
\altaffiltext{25}{Naval Research Lab., 4555 Overlook Av., SW Washington,
DC 20375-5352}
\altaffiltext{26}{Fermi National Accelerator Laboratory, Kirk Road and Pine 
Street, Batavia, IL 60510}
\altaffiltext{27}{W. W. Hansen Experimental Physics Laboratory and Department
of Physics, Stanford University, Stanford, CA 94305}
\altaffiltext{28}{Department of Astronomy, University of Florida, Gainesville,
FL 32601}
\altaffiltext{29}{University of Colorado, JILA, Campus Box 440, Boulder,
CO 80309-0440}
\altaffiltext{30}{Pennsylvania State University, Dept. of Astronomy,
525 Davey Lab, University Park, PA 16802}
\altaffiltext{31}{Osservatorio Astronomico di Torino, Strada Osservatorio 20,
I-10025 Pino Torinese, Italy}
\altaffiltext{32}{Joint Astronomy Centre, 660 N. Aohoku Place, University
Park, Hilo, HI 96720}
\altaffiltext{33}{Copernicus Astronomical Center, Polish Academy of Science,
Warsaw, Poland}
\altaffiltext{34}{Tuorla Observatory, Tuorla 21500 Piikki\"o, Finland}
\altaffiltext{35}{NOAO/KPNO, N. Cherry Avenue, P.O. Box 26732, Tucson,
AZ 85726}
\altaffiltext{36}{Mets\"ahovi Radio Research Station, 02540 Kylmala, Finland}
\altaffiltext{37}{Osservatorio Astronomico, Universit\`a di Perugia, I-06100
Perugia, Italy}
\altaffiltext{38}{Department of Physics, University of Milan at Como, 
Via Lucini 3, I-22100 Como, Italy}
\altaffiltext{39}{MS 306-388, Jet Propulsion Laboratory, California Institute
of Technology, 4800 Oak Grove Drive, Pasadena, CA 91109}
\altaffiltext{40}{Pomona College, Department of Physics \& Astronomy,
610 College Ave., Claremont, CA 91711-6359}

\begin{abstract}

The blazar 3C~279, one of the brightest identified extragalactic objects in 
the $\gamma$-ray sky, underwent a large (factor of $\sim$10 in 
amplitude) flare in $\gamma$-rays towards the end of a 3-week pointing by 
CGRO, in 1996 January-February. The flare peak represents 
the highest $\gamma$-ray intensity ever recorded for this object. During the 
high state, extremely rapid $\gamma$-ray variability was seen, 
including an increase of a factor of 2.6 in $\sim$8 hr, which strengthens the
 case for relativistic beaming.  Coordinated multifrequency 
observations were carried out with RXTE, ASCA, ROSAT and IUE and from many 
ground-based observatories, covering most accessible 
wavelengths. The well-sampled, simultaneous RXTE light curve shows an 
outburst of lower amplitude (factor of $\simeq$3) well correlated 
with the $\gamma$-ray flare without any lag larger than the temporal 
resolution of $\sim$1 day.  The optical-UV light curves, which are not 
well sampled during the high energy flare, exhibit more modest variations 
(factor of $\sim$2) and a lower degree of correlation. The flux at 
millimetric wavelengths was near an historical maximum during the 
$\gamma$-ray flare peak and there is a suggestion of a correlated  decay.
We present simultaneous spectral energy distributions of 3C~279 prior to and
 near to the flare peak.  The $\gamma$-rays vary by more 
than the square of the observed IR-optical flux change, which poses some 
problems for specific blazar emission models.  The synchrotron-
self Compton model would require that the largest synchrotron variability 
occurred in the mostly unobserved sub-mm/far-infrared region.  
Alternatively, a large variation in the external photon field could occur 
over a time scale of few days. This occurs naturally in the ``mirror'' model, 
wherein the flaring region in the jet photoionizes nearby broad-emission-line
 clouds, which in turn provide soft external photons that 
are Comptonized to $\gamma$-ray energies.

\end{abstract}


\section{Introduction}

The remarkable emission of blazars in the MeV-GeV energy range, 
relativistically enhanced by Doppler beaming, has made them the only 
class of Active Galactic Nuclei detected by the EGRET instrument on CGRO
 (Thompson et al. 1995). The quasar 3C~279 ($z$ = 0.538), the 
first radio source in which superluminal motion was discovered, is a prototype
 of the blazar class. It is the second brightest $\gamma$-ray 
blazar (Kniffen et al. 1993), the brightest being PKS~1622-297;
(Mattox et al. 1997). 

Violent variability is a distinguishing property of
 blazars and the $\gamma$-ray emission is no exception, varying with 
large amplitude on time scales of days or less (see recent review by Hartman 
1996), implying a very compact emission region.
The radio to UV continuum from blazars is commonly interpreted as synchrotron
 radiation from high energy electrons in a relativistic jet, 
while the MeV-GeV photons are  believed to be emitted via inverse Compton 
scattering of soft seed photons by the same electrons (e.g., 
Ulrich, Maraschi, \& Urry 1997). 
Finding correlations among the variations at high (X- to
$\gamma$-ray) and low frequencies is therefore critical to understanding which
 ranges of the Compton and synchrotron components are 
due to the same electrons and to clarify the nature of the seed photons 
available for scattering, namely whether they are generated within 
the jet (synchrotron-self Compton, SSC) or in regions external to the jet, 
like the accretion disk or the broad-emission-line clouds (external 
Compton, EC).

Multiwavelength observations of blazars in conjunction with EGRET pointings
 have been obtained at several epochs. However, either the 
monitoring was too sparse or the source was not active during the campaign,
 so that detections of correlated multiwavelength variability on 
short time scales are tentative (3C~279, Maraschi et al. 1994; Hartman et al.
 1996; OJ~287, Webb et al. 1996; PKS~0537-441, Pian et al. 1997).
A comparison of the spectral energy distribution of 3C~279 during the 
historically brightest state of the source (1991 June) with the lowest 
state ever observed (1992 December - 1993 January) showed that the 
$\gamma$-ray flux variation between the two epochs was larger than 
at any other wavelength (Maraschi et al. 1994), as predicted qualitatively by
 the SSC model (Maraschi, Ghisellini, \& Celotti 1992).  
Multiwavelength variability between 1991 June and October  was found to 
follow the same behavior (Hartman et al. 1996).  The larger 
$\gamma$-ray variability could also be accommodated within an EC scenario 
provided there was a change in the bulk Lorentz factor of the 
$\gamma$-ray emitting plasma, or the external photon field varied for some 
other reason, possibly as a result of enhanced photoionization of 
surrounding broad-line clouds by the jet itself.

The multiwavelength campaign on 3C~279 in 1996 January-February was organized 
as a 2-week coordinated program of the CGRO, 
ROSAT, RXTE and IUE spacecraft, plus a 1-week Target of Opportunity extension 
triggered by the high intensity measured with EGRET 
during the first week. The aim was to follow the evolution of a 
short-time-scale outburst at all frequencies, in order to constrain the 
possibility 
of a variation of the bulk Lorentz factor. This would allow discriminating
 between the possible models, and clarifying the nature of the seed 
photons being inverse Compton-scattered to $\gamma$-ray energies.
The campaign also benefitted  from the simultaneous and quasi-simultaneous
 observations with the HST, ASCA, and ISO satellites, as well 
as with many ground-based radio and optical telescopes. The final data set 
is rich in both temporal and wavelength coverage to an extent 
unmatched by any other blazar.
In this paper we present the observations conducted at the 
radio-to-$\gamma$-ray facilities participating in the multiwavelength
monitoring (\S~2), the light curves obtained (\S~3), and the spectral energy 
distributions before and near the flare peak (\S~4). We then compare our 
results with those at previous epochs and discuss constraints on theoretical 
models (\S~5) and summarize our findings (\S~6). 

\section{Observations}

In the following we give the essential information on the observations at 
each wavelength and summarize the results in Table~1.  The 
multiwavelength light curves of 3C~279 from 1996 January 11 through 1996 
February 13 are shown in Figure 1 on a logarithmic scale. 
Included are data with the most complete temporal coverage and at the full 
range of wavelengths.  We defer to separate papers for a 
complete presentation and for details about data analysis. In particular, a 
complete account of the $\gamma$-, X-ray and ISO  observations, 
data reduction and analysis will be given in Hartman et al. (1997), McHardy 
et al. (1997), and Barr et al. (1997), respectively.

During the CGRO observations from 1996 January 16-30, 3C~279 was close to the
 center of the field of view of EGRET and COMPTEL 
(5.0-6.7 deg).  OSSE began observations on 1996 January 24.  Due to the 
outstanding brightness detected by EGRET, the pointing was 
extended through 1996 February 6 as part of a Target of Opportunity program.
The high emission state of 3C~279 made it possible to detect significant 
signal with EGRET for integration times of 1 day, and even of 8 
hours during the flare.  The light curves are shown in Figure 1a.
Analysis of the
EGRET spectrum during the flare, from February 4-6, yielded an energy index 
$\alpha_\nu = 0.97 \pm 0.07$ between 30 MeV and 10 GeV, 
and $\alpha_\nu = 1.07 \pm 0.09$ was found for the period January 16-30.  
We use the convention $f_\nu \propto \nu^{-\alpha_\nu}$.
COMPTEL detected the source at energies above 3 MeV. Over the whole period 
the average flux in the 10-30 MeV band was (2.6 $\pm$ 
0.6) $\times 10^{-5}$ photons s$^{-1}$ cm$^{-2}$.  The average energy 
spectrum tends to be hard ($\alpha_\nu < 1$), 
however, the power-law slope cannot be determined accurately due to 
non-detection below 3 MeV.
The source was in the OSSE field of view from January 24 to February 7, and 
was detected in each of the two weeks at a high confidence 
level.

The RXTE satellite began observing 3C~279 less than a month after launch and
 monitored the source for 20 minutes daily from  January 21 
to February 10 during its performance verification phase (McHardy et al.  
1997), preceded by 6 days of observations with the ROSAT-HRI 
(January 14-20) and accompanied by one 20-kilosecond ASCA pointing on  
January 27 (Makino et al. 1996).  The RXTE data were 
calibrated by performing background subtraction from slewing data; the ASCA
 spectral index $\alpha_\nu = 0.7$ was used to calculate flux 
densities. The robustness of the RXTE background modelling is demonstrated
 by the agreement of the flux densities at the low end of the 
RXTE energy range and the high end of the ROSAT energy range on the day(s) 
in which their coverage overlapped.  The X-ray light curve 
is shown in Figure 1b.

IUE observed 3C~279 at approximately daily intervals from January 15.6 to 
February 6.8 with the LWP camera and on one occasion 
(January 25) with the SWP. The 13 LWP spectra were reduced and calibrated
 according to the Final Archive processing routine which 
adopts the NEWSIPS method for spectral extraction (Nichols \& Linsky 1996).
 Ly$\alpha$ emission (1216 \AA) is clearly visible on the SWP 
spectrum redshifted to $\sim$1870 \AA\ with a dereddened intensity of 
(5 $\pm$ 2) $\times 10^{-14}$ erg s$^{-1}$ cm$^{-2}$.  No emission 
line is present in the LWP spectra.
The LWP spectral signal was integrated and averaged in the 2500-2700 \AA\ 
interval, where the camera sensitivity is highest and the solar 
scattered light contamination (which might have been present in the first 
half of the monitoring) is negligible. The SWP signal was averaged 
in the 1400-1600 \AA\ range, where the camera sensitivity is high and no 
emission lines are superposed on  the continuum. 
Uncertainties are computed as in Falomo et al. (1993).
The LWP light curve is shown in Figure 1c.

The source was observed with the HST Faint Object Spectrograph using the 
G130H and G190H gratings, exposed for 2820 and 2250 
seconds, respectively, just before the start of the multifrequency campaign 
on 1996 January 8 as part of a different program whose results 
will be reported elsewhere (Stocke et al. 1997).  The shape of the dereddened 
spectral flux distribution in the interval 1300-2240 \AA\ is 
described by a power-law with energy index $\alpha_\nu = 1.81 \pm 0.05$.
Although not obtained during the EGRET pointing, these data are of interest 
here since they yield a reliable measure of the Ly$\alpha$ 
intensity which is important in estimating the inverse Compton contribution 
from external seed photons.  The dereddened line intensity is (4 
$\pm$ 1) $\times 10^{-14}$ erg s$^{-1}$ cm$^{-2}$. 

Optical BVRI photometry was obtained at several different sites listed in 
Table~2. The R-band has the best temporal coverage including one 
point close to the peak of the $\gamma$-ray flare, so only those data are
 shown in Figure 1d. The data in the B-band are very sparse; those 
in V- and I-bands show the same behavior as the R-band light curve within the 
uncertainties.  The conversion of optical magnitudes to fluxes 
has been done following Bessel (1979).  For a presentation of the complete 
data set of ground-based optical, near-IR, millimeter and radio 
observations, as well as for the IUE data related to this campaign, we defer
 to a separate paper.
The near-IR emission of 3C~279 was measured in the J, H and K bands at CTIO 
on January 31 and February 3. Only two data points were 
obtained within the time span of the campaign for each filter (Table 1). The 
conversion from JHK magnitudes to fluxes follows Bersanelli, 
Bouchet, and Falomo (1991).

The source was observed at millimeter and sub-millimeter wavelengths at the 
JCMT with both heterodyne and bolometer receivers as part of 
an extensive campaign that lasted through 1996 June. Few observations were 
obtained during the campaign reported here, but they were 
close in time to the $\gamma$-ray peak. The 0.45- and 0.8-mm data are shown 
in Figure 1e.  At longer wavelengths the variations were smaller.
Bolometric observations at millimeter wavelengths were carried out with the 
30m IRAM telescope using the IRAM/MPI 7-channel bolometer 
on 1996 January 13.  The nominal frequency of the bolometer is 250 GHz, the 
bandwidth about 60 GHz. The observations were carried out 
under poor weather conditions.  Observations of Uranus in the same night 
after weather conditions significantly improved were used for the 
flux calibration, assuming a flux of 35.18 Jy. The standard recommended 
gain-elevation correction was applied. The resulting fluxes of the 
two observations were (33.7 $\pm$ 0.3) Jy and (18.2 $\pm$ 1.1) Jy, where
 the errors are the rms of the single scans within each 
observations. We attribute the difference in the results to the changing
 weather conditions, and adopt a value of (26 $\pm$ 8) Jy.

Radio observations at 37 and 22 GHz were conducted at the Mets\"ahovi 
Radio Research Station from January 3 to February 11 and at 4.8, 
8 and 14.5 GHz from January 2 to March 1, altogether, at the University of 
Michigan Radio Astronomical Observatory, as part of longterm 
monitoring programs.  The resulting light curves at the three highest radio
 frequencies are shown in Figure 1f.

\section{Comparison of Multiwavelength Light Curves}

The 1-day binned EGRET light curve shows an extraordinary flare peaking on 
February 5 (Fig. 1a).  Before  January 30, the fluctuations 
visible to the eye in the $\gamma$-ray light curve are probably not due to 
real variability (the probability of variability is 30\%, according to a 
$\chi^2$ test).  The peak flux represents an increase by a factor of 10 with
 respect to the average level between January 20 and 30.  The 8-hour binned
EGRET light curve during the outburst appears modulated by high amplitude 
variations, the largest of which, a factor of 4-5 in 
one day, has a doubling time of only $\tau_D = 6$ hours 
($\tau_D \equiv {F_{initial}\over \Delta F} \cdot \Delta t$).
Between the January 16-30 and the Target-of-Opportunity periods,  the flux in
 the 10-30 MeV range (COMPTEL) increased by a factor of 
3.6 (2.5$\sigma$ significance  level).
No significant variability on timescales of days was found in the OSSE data, 
according to a $\chi^2$ test.  

The X-ray light curve also shows a large outburst, well correlated in time 
with the $\gamma$-ray flare but of lower amplitude (factor of 3, Fig. 
1b).  Any possible lag is less than the temporal resolution of 1 day, as 
confirmed by an analysis with the Discrete Correlation Function 
method (DCF, Edelson \& Krolik 1988).  The width of the outburst is about 7 
days in X-rays, where the data extend from  the pre-flare state 
to the decay, while the $\gamma$-ray coverage ends one day after the flare 
peak.  The ROSAT HRI data did not reveal any variability larger 
than 10\%, therefore the average 1 keV flux has been reported here as well as
 for the ASCA observation.  

The light curve at 2600 \AA\ (IUE-LWP) is reasonably well-sampled during the
 first part of the campaign but not toward the end, when the 
$\gamma$-ray flare occurred (Fig. 1c). It shows a broad minimum at $\sim$ 
January 25-26 followed by a rise of almost a factor of 2, but 
with a three-day gap before and up to the $\gamma$-ray peak. If the UV minimum
 were associated with the (possible) minimum in the 
EGRET light curve at January 28, this would indicate a correlation with the 
UV leading the $\gamma$-rays by $\sim$2.5 days. In this case 
the UV maximum would have occurred before the $\gamma$-ray peak, during the
 gap in IUE monitoring between February 1 and 5.  
Unfortunately, the UV and $\gamma$-ray light curves have too few points to
 apply the DCF method efficiently, so no robust result can be 
found from their cross-correlation.

The R-band light curve is similar to the UV light curve in showing a broad 
minimum on days  January 26-28 followed by a rise (Fig. 1d).  
Again, the sampling around the $\gamma$-ray flare is very poor. 
One observation very close to the flare peak yields a flux higher than the 
average around day January 28 by a factor of 1.6.  The behavior in V and I 
(not shown) is similar.  On the whole, the optical light curves 
suggest that the minimum occurs later than the UV minimum by 1-2 days. 
 They resemble the $\gamma$- and X-ray light curves in the flare 
rise, but differ significantly in having values quite close to those at the
 peak also at other epochs (e.g., around January 20) when the high 
energy light curves have values much lower than the peak.  In other words,
 the flare stands out in the high energy light curves while it is not 
apparent as such in the UV-optical light curves.
A near-infrared flux increase was observed, whose amplitude is a factor of 
1.3 in J and H, and 1.2 in K band.  Therefore, within the limited 
sampling, the JHK data are consistent with the rising trend of the other 
light curves.

The submillimeter data are rather sparse, but show variability consistent 
with the occurrence of a flare around  February 3 (Fig.~1e).  The 
sparse sampling prevents us from determining conclusively that the 
submillimeter peak actually occurred on February 3; it could as well have 
occurred on January 30, 31 or February 1, 2.  Observations at 0.45, 0.8, 1.1,
 1.3 and 2 mm on February 5 and 6 show a decline in flux, of 
decreasing amplitude with increasing wavelength, corresponding to the 
$\gamma$- and X-ray decline after the outburst.  We notice that the 
level of the mm/sub-mm flux reached during the present campaign has been 
exceeded only once (in 1994) in the last 7 years, and in 1996 
May a further increase by 20-30\% was recorded.

At radio frequencies the variability is highly significant (Fig. 1f). There
 is a nearly monotonic increase of $\simlt$30\% amplitude at 37 GHz 
from January 18 to  February 9, and a smaller increase at lower frequencies.
  A 6-7 Jy rise in 20 days is rare in the 16-year Mets\"ahovi 
database.  The brightness reached its historic maximum (since 1980) during 
May-June 1996, a time delay of about 4 months relative to the 
X- and $\gamma$-ray flare.

\section{Radio-to-$\gamma$-ray Energy Distributions}

The multiwavelength data collected during the 1996 monitoring campaign allow
 us to follow the evolution of the overall spectrum of 3C~279 
from a quasi-stationary state through the development of a dramatic
 high-energy outburst. There is no unique definition of a pre-flare state. 
In Figures 2 and 3 we show average fluxes in the period January 24-28 (where
 available), which includes the UV and optical minima.
The epoch of the high energy outburst is well covered at most wavelengths, no
 more than 2 days from the $\gamma$-ray peak.  We can 
therefore construct a reliable spectral energy distribution (SED) for the
 highest state.  We averaged the available data in a 2-day window 
centered on the $\gamma$-ray peak (February 4-6).  The resulting SED for the
 flaring state is shown in Figures 2 and 3.
Near-IR, optical and UV data (Table 1) have been corrected (shown in Figures
 2 and 3) according to Cardelli, Clayton and Mathis (1989) for 
Galactic interstellar extinction using $N_H = 2.22 \times 10^{20}$ cm$^{-2}$ 
(Elvis et al. 1989), a gas-to-dust ratio $N_H/E_{B-V} = 5.2 
\times 10^{21}$ cm$^{-2}$ mag$^{-1}$ (Shull \& Van Steenberg 1985), and a
 total-to-selective extinction ratio $A_V/E_{B-V} = 3.1$ (Rieke 
\& Lebofsky 1985).  The X-ray count rates have been converted to flux units 
using a power-law energy index of 0.7 derived from the ASCA 
2-10 keV observations.  The $\gamma$-ray photon counts have been converted to 
fluxes at 0.4 GeV according to Thompson et al. (1996).

The spectrum consists of two broad humps with peaks at $\sim 10^{12}-10^{13}$ 
Hz and $ 10^{22}-10^{24}$ Hz.  ISO data (Barr et al. 1997) 
will be of great importance to determine the shape of the SED in the range 
where the maximum synchrotron power is expected to be emitted.  
It is interesting to note that the sub-millimeter spectral slope during the 
flare ($\alpha_\nu = 0.38 \pm 0.09$ on February 5 and $\alpha_\nu = 
0.51 \pm 0.08$ on February 6) is roughly the same as the hard-X-ray to 
MeV-$\gamma$-ray spectrum ($\alpha_\nu \sim 0.6$), as expected 
if the same electrons are responsible for the synchrotron and inverse 
Compton-scattered radiation at those energies.
Comparing the flare and pre-flare states it is clear that the high energy 
spectrum (X- to $\gamma$-rays) is harder at the flare peak, as 
implied by the larger amplitude of the $\gamma$-ray variation. 
From near-IR to UV frequencies the flare versus pre-flare variations are
smaller than in X- and $\gamma$-rays. Comparing simultaneous J, H and K fluxes
 at two epochs suggests again that the variability amplitude 
increases with frequency, but the effect does not show up comparing UV to V,
 R or I band variations.  There is little information on the pre-flare 
fluxes at still lower frequencies, except for the radio band which is
 only weakly coupled to the rest of the SED. We note, however, that 
from the few data points available the amplitude of the variations at 0.45 
and 0.8 mm is comparable to that of the simultaneous X-ray 
variations.

The SEDs of 3C 279 obtained during the 1991 June  high state and the 1993 
January  low state are also shown in Figure 3 for comparison 
with the flare and pre-flare SEDs derived here. The 1991 $\gamma$-ray data
 are averaged over the 2-week pointing which included the 
flare. The X-ray and R band observations were simultaneous, while the other
 measurements were close in time except for the UV spectrum 
which was obtained one month later (Hartman et al. 1996).

\section{Discussion}

In early 1996, the blazar 3C~279 was observed in its highest $\gamma$-ray 
emission state ever. The pre-flare flux level (before 1996 
February 1) was comparable to the average state in 1991 June (Kniffen et al.
  1993; Hartman et al. 1996).  The presently observed 
maximum exceeds by a factor of $\sim3$ the peak of the 1991 June 24--25 
outburst, the brightest state recorded previously, and is 
$\sim90$ times higher than the historical $\gamma$-ray minimum seen with
 EGRET in 1992 December--1993 January (Maraschi et al. 1994).
Inspection of Figure~1 indicates decreasing variability amplitude with 
decreasing energy (within either the synchrotron or inverse Compton 
component), which is a common characteristic of blazar variability (e.g., 
3C~279 itself, Maraschi et al. 1994; PKS~2155--304, Urry et al. 
1997).  In addition, the X-ray emission during the 1996 outburst was higher 
than measured in 1991 June with {\it Ginga} over approximately 
the same energy range (Fig.~2).  Thus, not only is the X-ray variability 
amplitude lower than the $\gamma$-ray during the 1996 flare, but 
over longer time-scales the overall amplitude is also lower.  Notice that 
the flaring multiwavelength SED in 1996 February presents an 
``inverted'' variation with respect to the 1991 state: while the 
$\gamma$-ray flux is {\it higher} than in 1991 by a factor of $\sim$4 and the 
optical-UV flux is {\it lower} than earlier by a factor of $\sim$1.5-2. 

During the 1996 observations, significant $\gamma$-ray variability was found 
on time scales comparable to the sampling resolution (i.e., 8 
hours). Such extremely fast variability has also been found in several other 
blazars (Hartman 1996; Mattox et al. 1997).  The amplitude and 
rapidity of these luminosity changes exceed a well-known limit based on 
accretion efficiency (Fabian 1979; Dermer \& Gehrels 1995), which 
probably occurs in blazars because their observed radiation comes from 
relativistically beamed jets (with unknown relation to accretion 
processes). 
The simultaneous variability in X- and GeV $\gamma$-rays shows for the first
 time that they are approximately co-spatial. This, plus the rapid 
$\gamma$-ray variability, gives a strong lower limit to the beaming factor 
from the condition that the emission region should be transparent to 
$\gamma$-rays ($\tau_{\gamma\gamma}\propto \delta^{-5} L/\Delta t$).  For the
 optical depth to photon-photon absorption to be less than 
unity, the required  beaming factor is $\delta_{\gamma}\ge 6.3$ or 
$\delta_{\gamma}\ge 8.5$ for photons of $\sim$1 or $\sim$10 GeV, 
respectively.  These values are derived following Dondi and Ghisellini 
(1995), but are somewhat larger than theirs due to the faster variability 
now observed.  An independent argument for relativistic bulk motion of the 
low-frequency emitting region comes from the limit to the X-ray 
flux produced by the self-Compton process (Marscher et al. 1979), which gives
 $\delta\ge 18$ (Ghisellini et al. 1993).  A third estimate 
comes from the observed superluminal expansion of VLBI-resolved knots, 
$\delta\sim$6 (preliminary estimate from Wehrle et al. 1997).

In low-frequency peaked blazars like 3C~279, high energy electrons in a 
relativistic jet radiate at radio through UV wavelengths via the 
synchrotron process, and can produce X- and $\gamma$-rays by scattering soft 
target photons present either in the jet (SSC) or in the 
surrounding «ambient" (EC, Maraschi, Ghisellini, \& Celotti 1992; 
Blandford 1993; 
Dermer, Schlickeiser, \& Mastichiadis 1992; Sikora, Begelman, \& Rees 1994). 
 The relative variability in the synchrotron and inverse 
Compton components can indicate the origin of these seed photons.  Specific,
 time dependent  models are clearly necessary for an in 
depth discussion but are beyond the scope of this paper. In the following we
 discuss in general terms different scenarios for the origin of the 
seed photons assuming that a single active «blob" in the jet is 
responsible for the variability.

The SSC model predicts that a change in the electron spectrum (intensity 
and/or shape) should cause larger variability in the inverse 
Compton emission than in the synchrotron emission because the energy 
densities of the seed photons and the scattering electrons vary in  phase. 
In a one-zone model, the peak flux of the inverse Compton SED should 
vary approximately quadratically with the peak flux of the synchrotron 
distribution (Ghisellini \& Maraschi 1996).
Between 1991 June  and 1993 January this quadratic variation condition was
 satisfied assuming the synchrotron peak was close to the IR 
band (Maraschi et al. 1994; Ghisellini \& Maraschi 1996), but for the 1996
 flare vs. pre-flare SEDs the amplitude of the $\gamma$-ray 
variation is {\it more than the square} of the IR-optical-UV flux variation.
  However, there are very few data close to the $\gamma$-ray 
maximum (the IR points are from February 3, which is at half maximum),
 and the synchrotron peak may also fall at lower frequencies ($\sim 
10^{13}$~Hz, as suggested by the strong flux at  millimetric wavelengths) 
where adequate variations could have occurred.  A further caveat 
is that different emission zones could contribute to the IR-optical flux, 
diluting the intrinsic variation due to the $\gamma$-ray emitting region.
We conclude that the SSC scenario can not be ruled out by the present data.

Alternatively, we consider the EC scenario, (Sikora, Begelman, and Rees 1994)
 where the seed photons are external to the jet and 
independent of it.  In this case: (i) the inverse Compton emission should 
vary linearly with the synchrotron emission for changes in the 
electron spectrum; and (ii) larger than linear variations of the inverse 
Compton emission can be explained if the bulk Lorentz factor of the 
emission region varies together with the electron spectrum. In the latter 
case the different beaming patterns of synchrotron and inverse 
Compton radiation should also be taken into account (Dermer 1995).  As for
 the SSC model, different emission zones contributing to the IR-
optical flux would dilute the intrinsic variation due to the $\gamma$-ray 
emitting region. 
While the second case is conceivable comparing SEDs separated by years, it 
is far less likely that the entire emission region could 
accelerate {\it and decelerate} significantly over the time-scale of the 
rapid flare observed here.
The first case is unlikely because of the apparent nonlinear response of 
the $\gamma$-rays to the synchrotron variability. 
 
An interesting alternative combining advantages of both the SSC and EC 
scenarios is the ``mirror'' model of Ghisellini \& Madau (1996).  
Here the seed photons are provided by rapidly varying broad-line emission 
from a few clouds close to the jet and photoionized by an active 
blob in it.  First, the photoionizing continuum is beamed and therefore 
intense and highly variable; second, the electrons in the jet see broad-
line emission from the nearest cloud(s) as beamed; and, third, the
 $\gamma$--ray emitting blob, approaching the clouds, will see an 
increasing radiation energy density due to the decreasing blob-cloud distance.
  These effects lead to a more-than-quadratic increase in 
$\gamma$-rays associated with variations in synchrotron emission from the
 active blob.  This picture requires rather special conditions in 
that the cloud(s) close to the jet must also have a large covering factor, 
to let their emission line flux  dominate the radiation energy density 
seen by the blob.
The observations presented here can be accounted for by the mirror model if
 the far-UV (photoionizing) emission varied during the flare by 
a factor of 3-4.  This was not directly observed but is consistent with an 
extrapolation from the UV variations.  
If an increase occurred as an active region of the jet approached one or more
 broad-line clouds lying within the jet's beam, the observed 
amplitude of $\gamma$-ray variability could be explained, at least 
qualitatively.  Also, the asymmetric shape of the X-ray curve, in which 
the decay is possibly faster than the rise, can be accommodated by the mirror
 model since the inverse Compton emission drops sharply 
(because of the narrow angular pattern of the beaming) once the active part
 of the jet passes the broad line cloud(s).

We note that no variations in the Ly$\alpha$ luminosity are seen in archival
 (IUE and HST) spectra of 3C~279, as opposed to a large 
historical variability of the continuum, implying that a steady component, 
like an accretion disk, rather than the jet beam, dominates the 
overall photoionization of the  broad line clouds (Koratkar et al. 1997). 
However the  jet could still play a significant role in powering the 
clouds close to it.  The mirror model could be tested in principle, even in
 the absence of any available $\gamma$-ray observations, by 
monitoring the Ly$\alpha$ emission line of 3C~279. 
A limited number of clouds, over a limited velocity range, should respond 
simultaneously to the most rapidly varying (time scales of days) jet 
emission. 
However, the observed variability amplitude may be small, being diluted by 
the overall broad line region emission.
The line intensity measurements of January 8 and 25 from the HST-FOS and
 IUE-SWP spectra respectively, indicating no change, are 
inconclusive because they both refer to the pre-flare epoch  and to similar
 continuum levels.  Moreover the IUE sensitivity is far too low to 
measure variations in the line profile.

\section {Summary}
Radio-to-$\gamma$-ray monitoring of the blazar 3C~279 in 1996 January-February
 recorded the highest $\gamma$-ray flux of the source 
ever measured.  A correlated flare at X- and $\gamma$-ray energies with an
 amplitude of a factor 3 and 10, respectively, is seen and 
completely resolved.  The data at optical and UV frequencies clearly show a 
flux increase correlated with the X- and $\gamma$-ray rise, 
although the poor sampling close to the flare peak prevents a precise
 measurement of the amplitude in these bands.
The millimetric flux measured only close to the flare peak shows variability
 which could be correlated with the high energy light curves.  The 
radio emission exhibited variations of remarkable rapidity and amplitude.
The relative amplitudes of the high energy and low energy light curves during
 the flare and the apparently stronger IR to UV emission in 1991 
June, when the average $\gamma$-ray flux was weaker, represent important
 challenges for our understanding of blazars. 
The data do not rule out SSC models especially if more than one zone 
contributes to the emission,  but are difficult to reconcile with a 
scenario in which the seed photons are provided by the ambient surrounding the
 jet and  independent from it.  
A picture in which the relativistic jet hits and ionizes a small fraction of
 the broad line clouds which then provide the photons to be inverse-Compton 
upscattered seems appealing and likely.  Sensitive measurements of
 variations in the profile of the strong Ly$\alpha$ line 
correlated with the beamed UV continuum could test this model.

\acknowledgements 

We are grateful to the staffs of ISO, IUE, HST, ASCA, ROSAT, XTE, CGRO, CTIO, 
JCMT, IRAM, IAC Tenerife, Mets\"ahovi Station, 
UMRAO, NURO, Lowell Observatory, Boltwood Observatory, Foggy Bottom
 Observatory, the Burrell Schmidt Observatory of Case Western 
Reserve University, Boltwood Observatory, the ESO NTT, the observatories of
 Torino, Tuorla, Perugia and the University of Florida.  We 
thank Charles Dermer for his comments on the paper.  A. E. Wehrle, S. C. 
Unwin, and W. Xu acknowledge support from the NASA Long 
Term Space Astrophysics Program.  E. Pian is grateful for hospitality at IPAC 
during development of this research and acknowledges 
support from NASA Long Term Space Astrophysics Program.  M. Urry, E. Pian, 
and J. Pesce acknowledge support from NASA grants 
NAG8-1037, NAG5-2538, and NAG5-3138.  J. Webb acknowledges support from NASA 
IUE Guest Observer Program.  M. F. Aller and H.D. 
Aller acknowledge support from NSF grant AST-9421979.  D. Backman, J. Dalton
 and G. Aldering were Visiting Astronomers at Cerro Tololo 
Interamerican Observatory, National Optical Astronomy Observatories, operated
 by the Association of Universities for Research in 
Astronomy, Inc. (AURA), under a cooperative agreement with the National
 Science Foundation.  The National Undergraduate Research 
Observatory (NURO) is operated by Lowell Observatory under an agreement with
 Northern Arizona University and the NURO Consortium.  
The observers from Franklin and Marshall College thank the University of 
Delaware / Bartol Research Institute's Space Grant Colleges 
consortium for partial support of NURO membership and observations.  The 
James Clerk Maxwell Telescope is operated by The Joint 
Astronomy Centre on behalf of the Particle Physics and Astronomy Research 
Council of the United Kingdom, the Netherlands Organisation 
for Scientific Research, and the National Research Council of Canada.  We 
thank C. Imhoff, N. Loiseau, and J. Nichols for assistance with 
IUE observations and data reduction, and E. Solano and W. Wamsteker for 
timely NEWSIPS reprocessing of part of the IUE spectra.

\newpage
% ---------------------    REFERENCES    ------------------------

%  --------------------------   FIGURES   ------------------------------
\newpage

\cl{\bf Figure Captions}

\figcaption{Multiwavelength light curves of 3C~279 during the EGRET campaign 
(1996 January 16 - February 6): {\it (a)} EGRET fluxes at $>$100 MeV binned
within 1 day (open squares) and 8 hours (filled squares) (referred to 400 MeV,
following Thompson et al. 1996); {\it (b)} X-ray fluxes at 2 keV: besides the 
RXTE data (open squares), the isolated ASCA (filled square) and ROSAT-HRI 
(cross) points are reported with horizontal bars indicating the total duration
of the observation; {\it ©} IUE-LWP fluxes at 2600 \AA; {\it (d)} 
ground-based optical data from various ground-based telescopes in the R band;
{\it (e)} JCMT photometry at 0.8 mm (open squares) and 0.45 mm (filled 
squares); {\it (f)} radio data from Mets\"ahovi at 37 GHz (open squares) and 
22 GHz (filled squares), and from UMRAO at 14.5 GHz (crosses).  Errors, 
representing 1-$\sigma$ uncertainties, have been reported only when they are 
bigger than the symbol size.}

\figcaption{Radio-to-$\gamma$-ray energy distribution of 3C~279 in low (open 
circles) and flaring state (filled circles) in 1996 January-February.  The 
data plotted correspond to the entries of Table~1, except that the UV, optical
and near-IR data have been corrected for Galactic extinction (see text).  The 
slope of the ASCA spectrum ($\alpha_\nu = 0.7$) has been reported normalized 
to the RXTE point closest in time.  The EGRET best fit power-law spectra 
referring to the January 16-30 (low state) and February 4-6 periods are shown, 
normalized at 0.4 GeV. Errors have been reported only when they are bigger 
than the symbol size.}

\figcaption{Same as in Fig. 2.
%Radio-to-$\gamma$-ray energy distribution of 3C~279 in low (open 
%circles) and flaring state (filled circles) in 1996 January-February.  The 
%data plotted correspond to the entries of Table~1, except that the UV, optical
%and near-IR data have been corrected for Galactic extinction (see text).  The 
%slope of the ASCA spectrum ($\alpha_\nu = 0.7$) has been reported normalized 
%to the RXTE point closest in time.  The EGRET best fit power-law spectra 
%referring to the January 16-30 (low state) and February 4-6 periods are shown,
%normalized at 0.4 GeV.  
For comparison, the SEDs in  1991 June (stars) and in
1992 December -  93 January (diamonds) are also shown (see Maraschi et al. 
1994).  Errors have been reported only when they are bigger than the symbol 
size.}

\end{document}